\pdfoutput=1
\documentclass[twocolumn,
 reprint,
superscriptaddress,
nofootinbib,
aps,
pra]{revtex4-2}

\usepackage[utf8]{inputenc}

\usepackage{graphicx} 
\usepackage{float}

\usepackage{physics}

\usepackage{listings} 
\usepackage{color}

\begin{document}

\setlength{\parskip}{0pt}

\title{Predicting the minimum control time of quantum protocols with artificial neural networks} 

\author{Sofia Sevitz}
\email[Corresponding author:]{\,sofiasevitz01@gmail.com}
\affiliation{Departamento de F\'{i}sica “J. J. Giambiagi” and IFIBA, FCEyN, Universidad de Buenos Aires, 1428 Buenos Aires, Argentina
}
\author{Nicolás Mirkin}
\affiliation{Departamento de F\'{i}sica “J. J. Giambiagi” and IFIBA, FCEyN, Universidad de Buenos Aires, 1428 Buenos Aires, Argentina
}
\author{Diego A. Wisniacki}
\affiliation{Departamento de F\'{i}sica “J. J. Giambiagi” and IFIBA, FCEyN, Universidad de Buenos Aires, 1428 Buenos Aires, Argentina
}

\date{\today} 

\begin{abstract}

Quantum control relies on the driving of quantum states without the loss of coherence, thus the leakage of quantum properties onto the environment over time is a fundamental challenge. One work-around is to implement fast protocols, hence the Minimal Control Time (MCT) is of upmost importance. Here, we employ a machine learning network in order to estimate the MCT in a state transfer protocol. An unsupervised learning approach is considered by using a combination of an autoencoder network with the k-means clustering tool. The Landau-Zener (LZ) Hamiltonian is analyzed given that it has an analytical MCT and a distinctive topology change in the control landscape when the total evolution time is either under or over the MCT. We obtain that the network is able to not only produce an estimation of the MCT but also gains an understanding of the landscape's topologies. Similar results are found for the generalized LZ Hamiltonian while limitations to our very simple architecture were encountered. 

\end{abstract}

\maketitle

\section{Introduction}

The current century promises an advent of revolutionary technologies based on quantum information and quantum computing \cite{nielsen_chuang_2010,QuantumComputingMotivation}. Quantum control is a crucial part in the advances of these fields given that it dictates how a system should be manipulated in order to achieve a target goal \cite{QOCEnExperimentos}. A possible approach is by applying optimal control techniques \cite{ReviewQOC,RabitzOrigin}. This way the problem is reduced to finding a time dependent control field $\epsilon$(t) that maximizes a certain objective function, J[$\epsilon$(t)]. This must be done while preserving the coherence of the system, thus the leakage of quantum information onto the environment over time is a fundamental challenge. An efficient way to work-around the difficulty is to implement rapid controls, that is to say, to maximize J[$\epsilon$(t)] in the fastest time possible \cite{introductionQC_Book}.

The speed of these protocols is bounded by the minimum evolution time needed for a system to reach a target state driven by a time-dependent control field, namely Minimun Control Time (MCT), not to get confused with the Quantum Speed Limit which is an upper bound on evolution times given by the Heisenberg time–energy uncertainty relation \cite{QSLPoggi, QSLNONMARKOV, QSLPhysicalProcesses, Deffner_2017}. While conceptually easy, in practice it is a whole other story. Having an estimation of the MCT is of uttermost importance given that it governs the success of a control protocol in the presence of the detrimental effects from the environment.

Many efforts have been made in developing optimal control methods, especially from a numerical approach. In this framework, the studies of quantum control landscapes are a key factor. They are essentially the mapping of the objective function J[$\epsilon$] on the field $\epsilon$ parameterized in time \cite{RabitzLandscapes}. Their topologies are not trivial, for instance, local maximums or traps appear whenever the control problem holds a constraint \cite{ConstrainedLandscapes}. On top of that, usually, optimal controls require very high parametrization dimensions that produce landscapes beyond our intuitive understanding  \cite{Parametrizaciones}. Thus, extracting useful information from them might be an unimaginable task. 

Over the past years, various optimization algorithms have been developed such a as Gradient-Ascent Pulse Engineering Algorithm (GRAPE) and Chopped RAndom-Basis quantum optimization (CRAB), see review in Ref. \cite{ReviewQOC}. Recently, machine learning has been studied as an alternate optimization tool \cite{GIANNELLI2022128054, MLPresentacionQCC,QCbasedML,Nature_QC_ML}. For example, in Ref. \cite{BurkovRL} a reinforcement learning algorithm has been proposed in order to find an optimal driving protocol for a state transition scheme. Another case is Ref. \cite{MLQuantumControlCNN}, where a supervised learning algorithm classifies randomness of a system in order to find an optima control policy.

In this paper we use machine learning but not as an alternate optimization tool. We instead aspire to answer if an Artificial Neural Network (ANN) is able to estimate the MCT just by training over the landscape of a system. On that note, we are also interested in the exploration if an ANN actually understands the topology of the landscape or merely detects that an extremity is reached. For this task, an unsupervised ANN scheme is selected in order to avoid any external bias, thus any information obtained from the network is purely due to the information stored within the dataset. An autoencoder extracts the important features of the dataset by means of dimension reduction that are later classified with the use of the k-means clustering tool. As a first approach, the toy model Landau-Zener (LZ) Hamiltonian is analysed given that it has an analytical MCT and a distinctive change in the landscape's topologies when the total evolution time is either under or over the MCT. We obtain that the ANN is not only able to predict the MCT but also gains an understanding of the landscape's topologies. We moved on to implement the network in the generalized LZ Hamiltonian where similar results are found beyond some limitations encountered within our simple scheme.

The paper is organized as follows. In Section II we present the models used in this work, LZ Hamiltonian and its generalization, while giving an overview of the optimal control framework. The unsupervised neural network approach implemented is detailed in Section III. The results obtained for the Landau-Zener Hamiltonian and its generalization are given in Section IV. Finally, in Section V we communicate the conclusions and aspirations for future works.

\section{Optimal Control framework and Models} \label{controlframework}

Quantum control theory is the theoretical framework that studies the manipulation of quantum systems. A typical driving Hamiltonian is given in the following way,
\begin{equation}
    \text{H}[\epsilon(t)]=\text{H}_0+\epsilon(t)\text{H}_c,
    \label{ControlHamiltonian}
\end{equation}
where $\text{H}_0$ is the isolated system one wants to manipulate, known as drift Hamiltonian, $\text{H}_c$ the control Hamiltonian and $\epsilon(t)$ a time-dependent control field parameterized in time \cite{laroccaLZ}. The evolution of the system is governed by the shaping of said field, also referred to as protocol. In this work, we consider the state transfer scheme; we seek to achieve a desired target state $\ket{f}$ from an initial state $\ket{i}$ after a given evolution time T. The protocol duration is divided into N$_{ts}$ uniform intervals of $\Delta t= \text{T}/\text{N}_{ts}$ and a piecewise function is used to parameterize the control,

\begin{equation}
    \epsilon(t)=
    \begin{cases}
        \epsilon_1 & \text{if } 0 < t \leq \Delta t\\
        \vdots \\
        \epsilon_{\text{N}_{ts}} & \text{if } (N_{ts}-1)\Delta t < t \leq \text{T}
    \end{cases}
\end{equation}

The measure used in this work to quantify how well a control protocol performs is the state fidelity, given by 
\begin{equation}
    \text{F} [\epsilon]=|\langle f |\text{U}_\epsilon(\text{T})| i \rangle|^2   \in (0,1).
    \label{Fidelity}
\end{equation}
Note that $\text{U}_\epsilon(T)$ is the evolution operator given by the Hamiltonian in equation (\ref{ControlHamiltonian}) at time T, thus it is dependent of the protocol. A simple view of equation (\ref{Fidelity}) discloses that F=1 corresponds to an optimal control, the achieved state after a given evolution time is equal to the desired one. 

In this work, we consider the paradigmatic Landau–Zener (LZ) Hamiltonian where the drift and control Hamiltonian are the following
\begin{equation}
    \text{H}_0=\frac{\delta}{2}\sigma_x \quad\text{and}\quad \text{H}_c=\sigma_z.
    \label{LZHamiltonian}
\end{equation}
Here $\sigma_x$ and $\sigma_z$ are the Pauli matrices, $\delta$ is referred to as the energy gap given that it measures the minimum separation between the energies of the system under no perturbation \cite{laroccaLZ}. From now one we use the natural units such that $\hbar=1$, thus the units of time T are [1/$\delta$]. This Hamiltonian is of particular interest within the field given that it describes a simple two level system that can be applied, for example, to a semiconductor quantum dot \cite{DiegoMurgidaTambo}. This model is a great starting point given that not only it has a closed solution for its minimum time \cite{Hegerfeldt} but also a rather simple landscape because that optima solutions are reached with just a two-dimensional parametrization, N$_{ts}=2$ \cite{laroccaLZ}.  

The typical transfer scheme is to take the initial state $\ket{0}$ and the desired final state $\ket{1}$, in the $\sigma_z$ base. \citet{Hegerfeldt} demonstrated that the MCT for the scheme is given by
\begin{equation}
    \text{T}_\text{MCT}=\frac{\pi}{\delta}.
\end{equation}

\par Taking advantage that we can visualize optimal landscapes of these systems, in this work we start off concentrating in the simple case N$_{ts}=2$. We parameterize the control field as a piecewise function in the following way

\begin{equation}
    \centering
    \epsilon(t)=
    \begin{cases}
        \epsilon_1 & \text{if } t \leq \text{T}/2\\
        \epsilon_2 & \text{if } t > \text{T}/2
    \end{cases}
\end{equation}

\begin{figure}[H]
    \centering
    \includegraphics[scale=0.3]{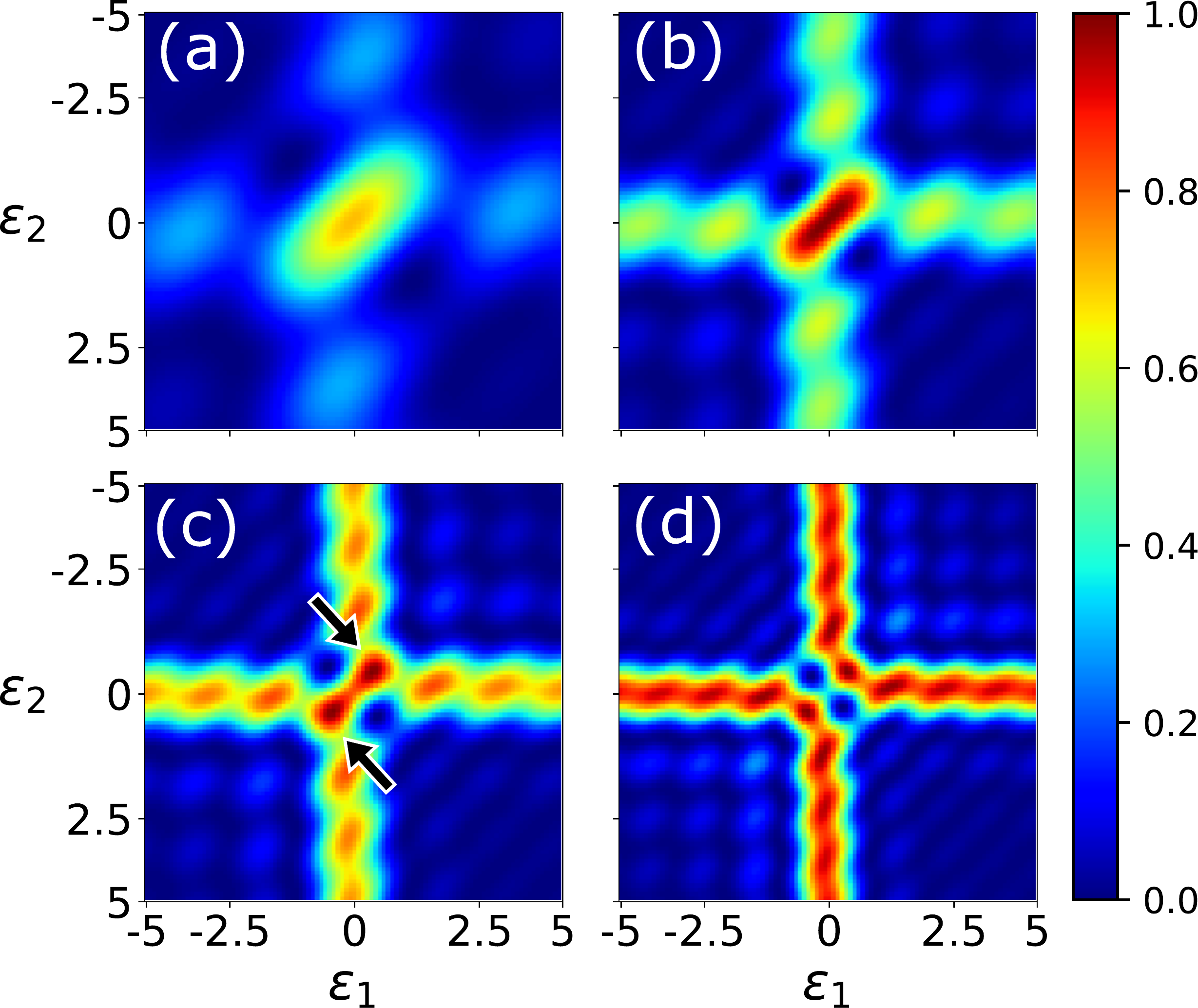}
    \caption{Control landscapes constructed with equation (\ref{Fidelity}) for the simple Landau-Zener Hamiltonian with $\delta=1$, thus the analytical MCT is T$_\text{MCT}$=$\pi$. Each panel corresponds to a different total evolution time: (a) T= 2, T= 3.14, (c) T=4 and (d) T= 5.}
    \label{LZ_landscapes}
\end{figure}

The controls ($\epsilon_1$ and $\epsilon_2$) are broken down into mesh points that conform the domain of the fidelity function forming the landscape. In Fig. \ref{LZ_landscapes} we show the mapping obtained for different evolution times: (a) T$<$T$_\text{MCT}$, (b) T$\simeq$T$_\text{MCT}$, (c) T$>$T$_\text{MCT}$ and (d) T$>>$T$_\text{MCT}$. 

\vspace{5mm} %vertical space

When T$<$T$_\text{MCT}$ [Fig. \ref{LZ_landscapes} (a)] the landscapes present a sub-optimal (F$<1$) single global maximum at the origin. Whereas, when the system has reached the MCT [Fig. \ref{LZ_landscapes} (b)], there is an optimal global maximum in the origin corresponding to F=1. This indicates that an optimal control is achieved when taking the protocol $\epsilon(\epsilon_1,\epsilon_2)=0$. Beyond this total evolution time, Fig. \ref{LZ_landscapes} (c), interesting and complex topologies appear. The center maximum is splitted into two symmetrical structures (highlighted with black arrows) that oscillate in time. Also, many sub-optimal maximums form along the vertical and horizontal direction that transform into optimal maximums over time [see Fig. \ref{LZ_landscapes} (d)]. What is important to note from this system is that there is a distinctive difference in the landscape's topologies with T$<$T$_\text{MCT}$ and T$>$T$_\text{MCT}$. 

The model of LZ can be generalized to an N-level system studied in Ref. \cite{AvoidedCrossings}. In this work, we consider the case where N=3 in which the driving Hamiltonian in the base $\{\ket{0},\ket{1},\ket{2}\}$ can be written as 

\begin{gather*}
    \text{H}_0=\frac{1}{2}\begin{pmatrix}
                0 & \Delta_A & 0\\
               \Delta_A & 0 & \Delta_B\\
                0 & \Delta_B & -2\delta
               \end{pmatrix} \\ \\ 
    \text{H}_c =\begin{pmatrix}
            1 & 0 & 0\\
            0 & 0 & 0\\
            0 & 0 & 1
            \end{pmatrix}.
\end{gather*}

This Hamiltonian is commonly used in in quantum optics given that it is suitable for describing a three-level system that leads to a STIRAP protocol when $\Delta_A$ and $\Delta_B$ are the controls \cite{STIRAPReview}. However, in this work we take $\Delta_A$, $\Delta_B$ and $\delta$ as fixed parameters. Under these conditions, the energy spectrum as a function of the control holds two avoided crossings, as opposed to the traditional LZ Hamiltonian that only has one. Although there is no closed analytical MCT solution in this case, we are able to obtain an empirical estimation from the control landscapes by taking the total evolution time of the landscape that reaches the highest fidelity within the dataset. This way, for $\delta=1$, we define an empirical MCT, T$_\text{MCT}$=5.31.

\begin{figure}[H]
    \centering
    \includegraphics[scale=0.3]{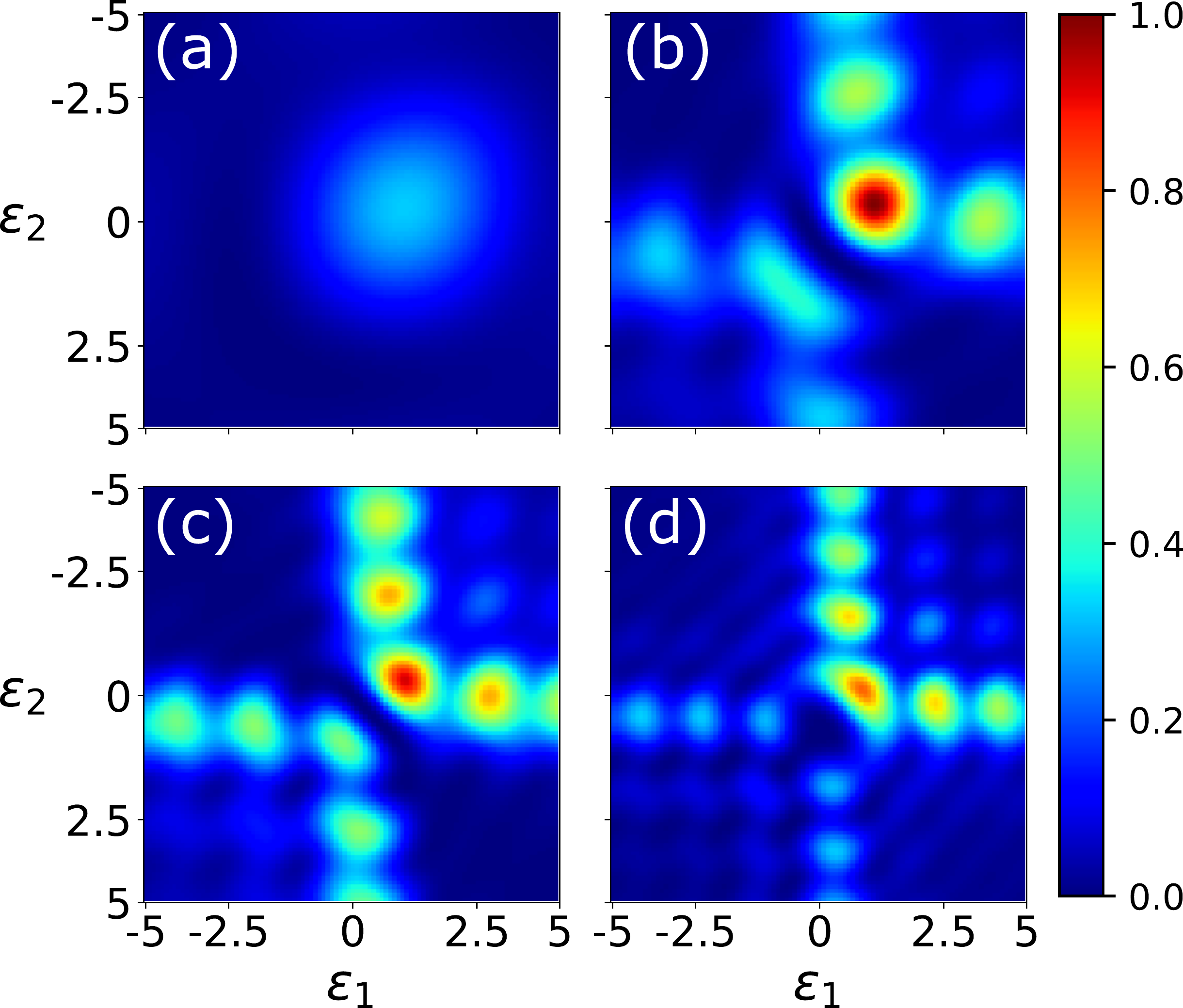}
    \caption{Landscapes for the generalized LZ Hamiltonian with N=3 and $\delta=\Delta_A=\Delta_B=1$. We take the total evolution times T as (a) T= 2.5, (b) T= 5.31, (c)  T= 7 and (d) T= 9. } 
    \label{fig:Landscapes_AC}
\end{figure}

The landscape topologies are very different for this system [Fig. \ref{fig:Landscapes_AC}]. In particular, they do not hold symmetric topologies and the global maximum is shifted as a function of $\delta$. More importantly, unlike LZ, this system does not exhibit a distinctive change in the landscape for when T$<$T$_\text{MCT}$ and T$>$T$_\text{MCT}$. For instance when T$>>$T$_\text{MCT}$, the region of the landscape we are exploring does not hold a optimal control protocol. Thus, in this sense, we believe that this is a good system to test the generalization properties of our proposed artificial neural network. 

\section{The unsupervised artificial neural network method}

\begin{center}
    \begin{figure}[H]
        \centering
        \includegraphics[scale=0.7]{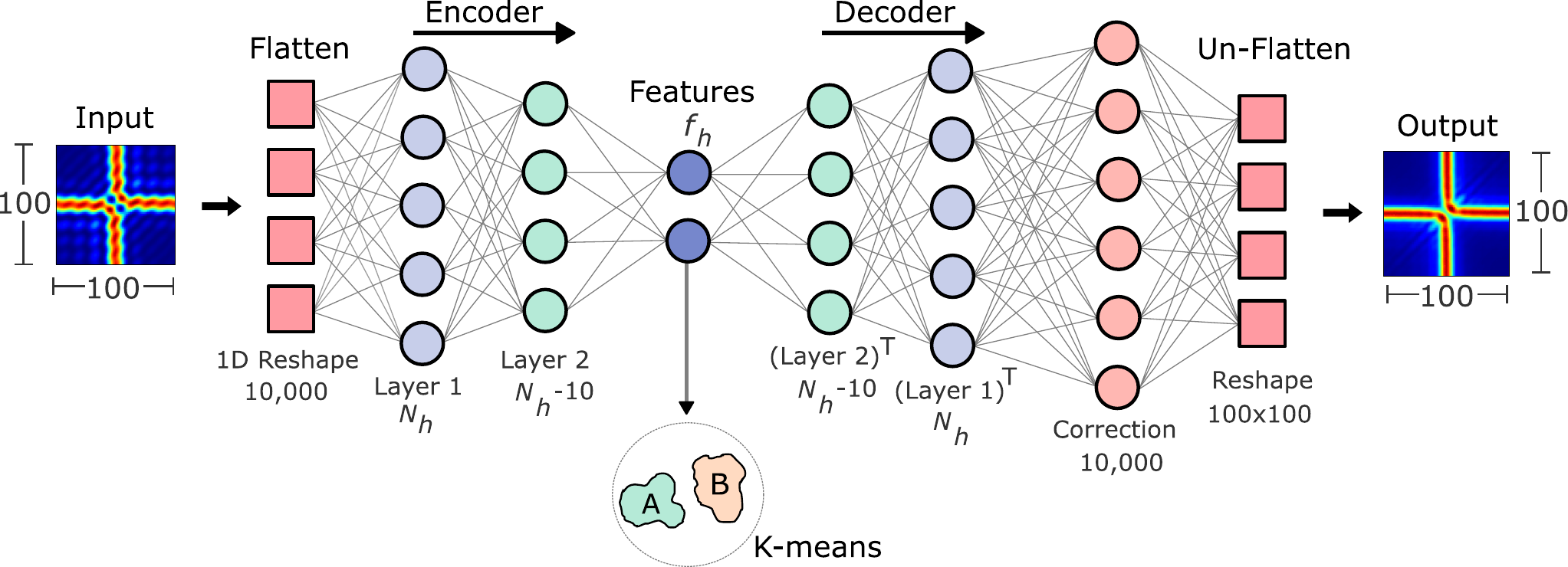}
        \caption{Architecture of the ANN used in this work. It is composed of two parts: the autoencoder network and the k-means clustering method. The autoencoder is composed of a flattened stage, encoder hidden layers, a layer with the smallest amount of nodes corresponding to the features, decoder hidden layers and finally a last layer that recovers the dimensions needed for the un-flattened to have the same dimensions as the input. Below each component a detail of the dimensions are given, where $N_h$ and $f_h$ describe the number of hidden nodes in the first hidden layer and the feature layer, respectively.}
        \label{fig:ANNArchitecture}
    \end{figure}
\end{center}

In this work, we would like to explore if an ANN can estimate the MCT by understanding a topology change within a landscape or simply detects that a pixel of the landscape obtains the value of $\approx 1$. Given that we do not want to add any external bias to the network and we usually don't know which is the real value of the MCT, we opt for a fully unsupervised approach by using a combination of an autoencoder network with k-means clustering. The architecture is illustrated in the Fig. \ref{fig:ANNArchitecture}. 

The autoencoder is a typical data reduction network. Its main task is to optimize the network such that the input layer is the same as the output layer while generating an effective representation of the data, commonly called as features. Given that the network evaluates the loss function comparing its output to its input, the training is done in an unsupervised or self-supervised fashion. The unsupervised k-means clustering method is then implemented over these extracted features and, by means of competition clustering, obtains a \textit{natural} grouping of data. Further details of the network are given in Appendix \ref{AppendixANN}.

The input dataset is composed of a collection of control landscapes with different evolution times T. Given that the first layer of the network flattens the dataset into a 1-dimensional array, this enables us to generalize the architecture for any parameterization dimension (specifically N$_{ts}>2$). The input data is randomly shuffled and separated using the holdout method, 70\% of the data is used for training while the remaining 30\% is reserved for validation. In order to avoid a bias training in the autoencoder and k-means, the training datasets are equally separated into two parts: autoencoder training dataset and k-means training dataset. For the same reason, the separation is also done for the validation dataset into autoencoder validation dataset and performance dataset. See Appendix \ref{AppendixANN} for further information.

\onecolumngrid
\twocolumngrid

The data manipulation consists of three steps: (1) autoencoder training, (2) k-means unsupervised clustering training and (3) the performance measure. In the following, we will break down each of these steps. In step (1), the autoencoder is trained using backpropagation with the classical Mean Squared Error (MSE) loss function. The training parameters are detailed in the Appendix \ref{AppendixANN}. Once the autoencoder is optimized, we move on to step (2). The features of both the cluster training dataset and the performance dataset are extracted. The cluster training features are then used to train the k-means model. In this step we implement the elbow method to optimise the number of clusters and obtained two clusters, ideally the classification of the landscapes corresponding to T$<$T$_\text{MCT}$ and T$>$T$_\text{MCT}$ (see Appendix \ref{AppendixANN}). The performance features are to be used in step (3) in order to evaluate an overall performance of the network. Given that we are taking an unsupervised approach there are no \textit{correct labels} to evaluate a performance measure, thus we appeal to an alteration of the confusion scheme presented in Ref. \cite{ConfusionScheme}. If the dataset one is trying to classify into two groups depends on a parameter that lies within a closed range, in our case the total evolution time T, then one can assume that there exists a critical point T$^\prime$ (within this range) that optimally classifies the dataset into the two groups. In order to find this T$^\prime$, one can take an auxiliary critical point T$_\text{aux}$ and create an auxiliary labeling of all data with parameters smaller than T$_\text{aux}$ with label A and the others with label B. Next, by varying T$_\text{aux}$, the performance curve is analysed. In this work the performance measure is the well used accuracy score. Although not the same scheme as presented in Ref. \cite{ConfusionScheme} we can still observe that the performance function with respect to the T$_\text{aux}$ has an universal shape, with a maximum at the correct critical point T$^\prime$. 

Ensemble averaging is implemented to smooth the output performance function. This method consists in combining many networks in order to form a collection of networks. The ensembles are composed of different combinations of number of nodes in the first hidden layer $N_h \in $[100, 110, 120,..., 190] and number of features extracted $f_h\in$[10, 20,.., 40], this way we take a total of 40 architectures. The performances of each of the architectures are later averaged in order to obtain a single mean clustering accuracy.

\section{Estimation of the minimal control time}

In this section we analyse the application of an ANN framework over the LZ Hamiltonian and its generalization presented in Sec. II. We commence with the landscapes from the LZ Hamiltonian, eq. (\ref{LZHamiltonian}),  with the parametrization N$_{ts}=2$ and $\epsilon_{\{1,2\}} \in [-5,5]$ with 100 points (see Fig. \ref{LZ_landscapes} for reference). We chose a sweep of total evolution times T from 0.01 to 10 in steps of 0.01, which yields a dataset of 1,000 landscapes (see Appendix \ref{AppendixANN} for more details). 

In Fig. \ref{fig:AccDelta1} we show the smoothed output performance function taking the energy gap $\delta=1$. The critical point is located slightly below the analytical MCT, with an estimated value of T$^\prime=2.9$ where the analytical MCT is T$_\text{MCT}=\pi$. In the inset of Fig. \ref{fig:AccDelta1} we show the ANN estimation as a function of the energy gap $\delta$. As we can see, the unsupervised network gives a good prediction of the MCT, that is to say, without any external guidance regarding the total evolution time of the landscapes, the network is able to identify a change within the time shuffled landscapes. We point out that in the main panel a minimum is formed near T$_\text{aux}=8.1$, further investigation is given in the Appendix \ref{LongTimes}. 

\begin{figure}[H]
    \centering
    \includegraphics[scale=0.5]{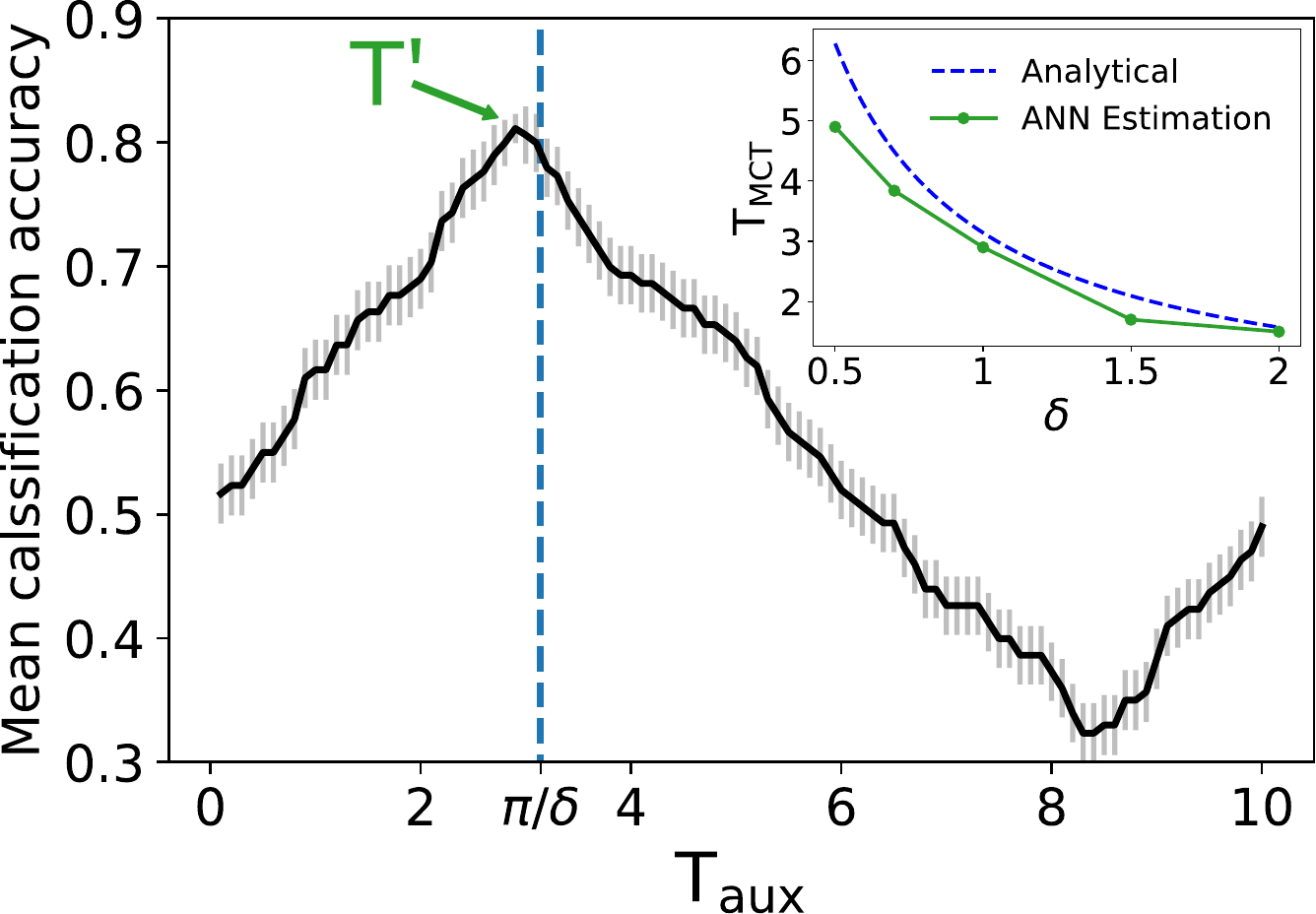}
    \caption{Ensemble averaged accuracy performance function for the Landau-Zener Hamiltonian with $\delta=1$. The minimal control time predicted by the unsupervised network, marked with T$^\prime$ is a good estimation with respect to the analytical value (dashed blue line). The predicted minimal control time for different values of $\delta$ are shown in the inset, overall it yields good estimations.}
    \label{fig:AccDelta1}
\end{figure}

\begin{figure}[H]
    \centering
    \includegraphics[scale=0.5]{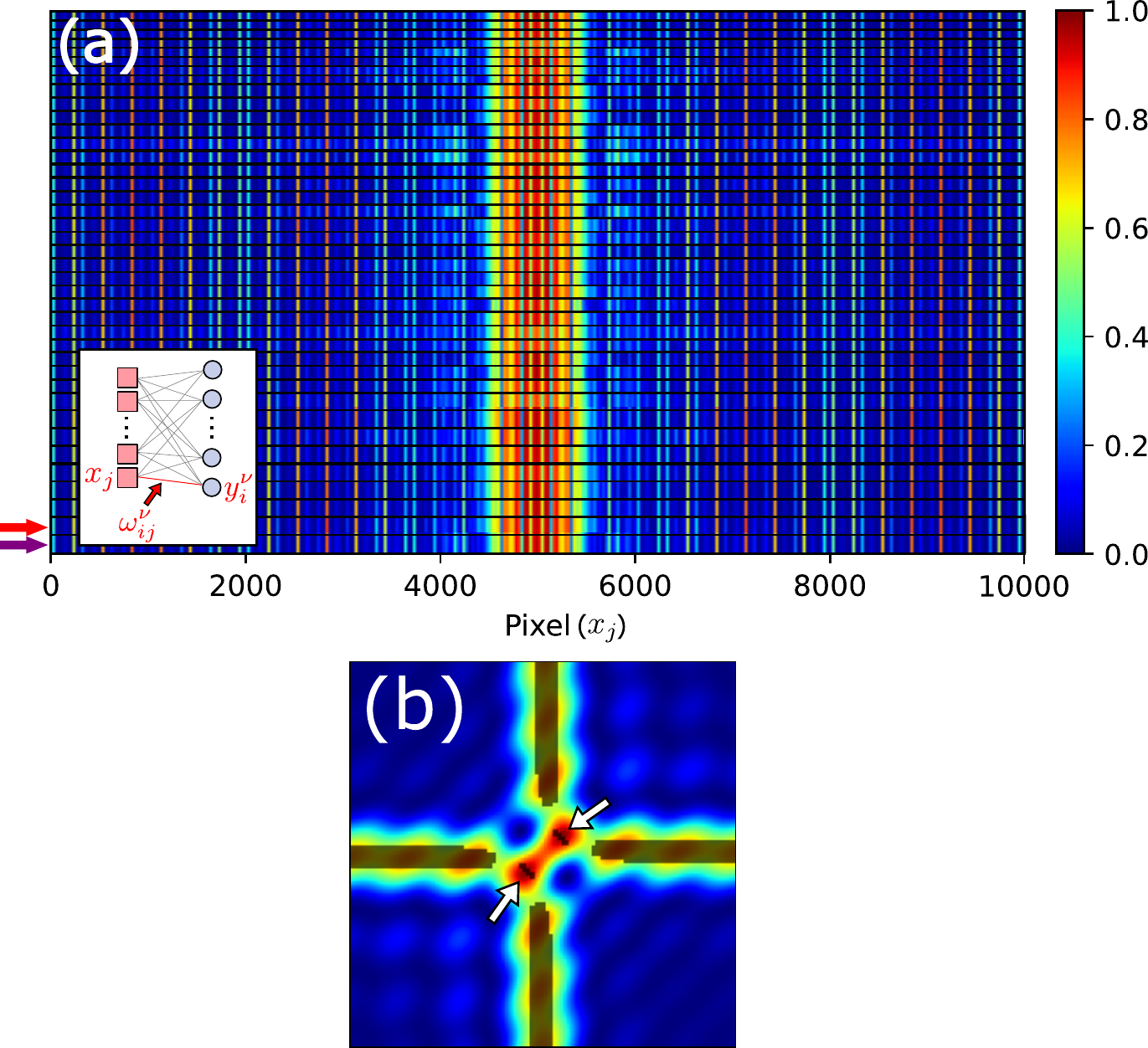}
    \caption{ Weight analysis of the network. (a) The normalized absolute values of the weights for all the nodes in each architecture against the pixels in the flattened landscape. (b) The pixels with high importance to the network (explained in the main text) highlighted over the landscape in black.}
    \label{fig:Weights}
\end{figure}

In the following we focus on understanding the underlining knowledge that the ANN  gains over the datasets. We analyse the weights between the nodes in the first layer of the ANN and the pixels in the landscapes (see inset of Fig. \ref{fig:Weights} (a) for reference). As mentioned in Sec. III, the collection of networks for ensemble averaging is composed of the different combinations of number of nodes in the first hidden layer and number of features extracted. Each one of these combinations form a specific architecture $\nu$. Given an activation function $\varphi$, the output function of each node $y_i^\nu$ for a given architecture $\nu$ can be written as
\begin{equation}
    y_i^\nu=\varphi\left(\sum_j \omega_{ij}^\nu x_j\right),
    \label{ANN_nodes}
\end{equation}
where $x_j$ is the pixel in the position j with the landscape reshaped into a 1-dimensional array and the weight between the input $x_j$ and the node $y_i^\nu$ is represented by $\omega_{ij}^\nu$. From eq. (\ref{ANN_nodes}) one can interpret the absolute value of each weight (i.e., $\abs{\omega_{ij}^\nu}$) as a measure of strength of the connection between $x_j$ and $y_i^\nu$. Thus, the most important pixels for the network correspond to those that have a higher $\abs{\omega_{ij}^\nu}$ compared to the rest. This interpretation is similar to the understanding of filters in convolutional neural networks. 

Fig. \ref{fig:Weights}(a) is composed of 40 subplots stacked on top of each other (separated by black horizontal lines) corresponding to each architecture. Each subplot shows the absolute values of the weights for all the nodes against the pixels in the flattened landscape. For instance, the first subplot (marked with a purple arrow) contains the 190 node weights corresponding to the architecture with 40 features. The second subplot (red arrow) corresponds to the architecture with also 190 nodes but 30 features. This way all the 40 architectures are stack up on top of each other.  In order to compare between architectures, we normalize over the maximum absolute weight of each architecture, $\displaystyle \left|w_{ij}^\nu \right|/\max_{kl} \left|w_{kl}^\nu \right|$. As explained, this plot gives an idea of the strength of the connection between the node $i$ and the pixel $j$. The network shows a symmetric behaviour with respect to pixel $x_j=$5000. This symmetry is attributed to the symmetrical aspect of the landscapes of the LZ Hamiltonian (see Fig \ref{LZ_landscapes}). In the following, we select all the pixels $x_j$ that satisfy the following condition: $\left\{x_j | \frac{1}{N} \sum_{i,\nu} \left( \left|w_{ij}^\nu \right|/\max_{ij} \left|w_{ij}^\nu \right|\right) \geq 0.7 \right\}$, where N is the total amount of architectures. This selection identifies the regions of the landscapes where the network is paying the most attention. As an illustration, in Fig. \ref{fig:Weights} (b) we take the landscape at T=4 and highlight in black the region of interest to the network. The ANN detects the symmetrical structures that form when the optimal maximum is separated into two (white arrows in Fig. \ref{fig:Weights} (b)). The network prioritize the horizontal and vertical direction corresponding to the sub-optimal maximums that form after the total evolution time T reaches the MCT. This indicates that the ANN also understands that a change in the topology of the landscape corresponds to a change in an underlying parameter of the system, in this case T. It is an exciting result given that this ANN can give insights of landscapes that are difficult to analyze visually, for instance, taking landscapes with N$_{ts}>2$. With this simple architecture we tested the LZ Hamiltonian with a N$_{ts}=3$ parametrization. Given that it is computationally demanding to construct the landscapes the larger the N$_{ts}$, we took a broad discretization of the mesh points; $\epsilon_{\{1,2\}} \in [-5,5]$ with 100 points and $\epsilon_3 \in [-5,-4,...,5,6]$ with 11 points. This yielded a predicted T$^\prime=4.34$, which we believe can be improved by taking a finer mesh discretization or adding complexity to our ANN. We note that this analysis was only possible because of the flatten stage of our ANN scheme, it would not have been possible if we chose, for instance, the widely used 2 dimensional convolutional neural networks.

\vspace{5mm} %vertical space

\begin{figure}[H]
    \centering
    \includegraphics[scale=0.3]{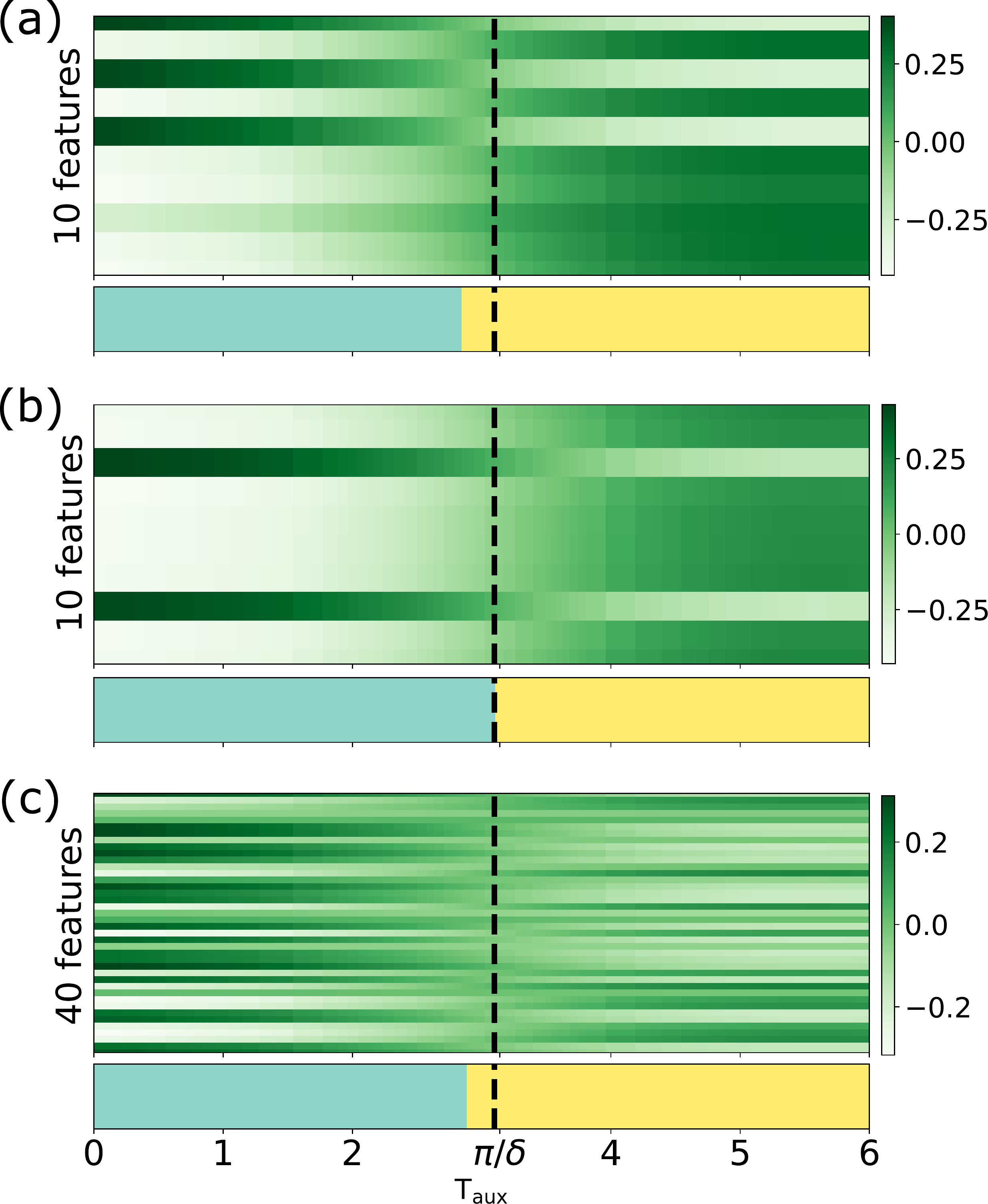}
    \caption{Features obtained from different architectures (a): 10 features and 100 hidden initial nodes, (b): 10 features and 120 hidden initial nodes and (c) 40 features and 100 hidden initial nodes.  Underneath each feature combination, the corresponding cluster assigned by the k-means clustering method with blue: group A and yellow: group B.}
    \label{fig:Features}
\end{figure}

\vspace{5mm} %vertical space

Next we analyse the features obtained from different architectures with their corresponding cluster assigned by the unsupervised k-means clustering. Fig \ref{fig:Features} (a) and (b) exhibit the 10 features assigned to each time for two different architectures: 100 and 120 initial hidden nodes, respectively. As seen, the features extracted from the landscapes corresponding to a total evolution time T $<$ T$_\text{MCT}$ are very different from those corresponding to T$>$ T$_\text{MCT}$. Near the theoretical T$_\text{MCT}$ (black dashed line) the features suffer a transition. We remind the reader that this is obtained with an unsupervised network and the landscapes are presented to the network in a shuffled time-order. This indicates that although we are not using feedback connections (typically used for time-dependent problems) our network still understands an underlining time order. The same was observed for all the architecture configurations, for instance, in Fig. \ref{fig:Features} (c) we exhibit the case with 40 features and 100 hidden nodes in the initial layer.

\begin{figure}[H]
    \centering
    \includegraphics[scale=0.3]{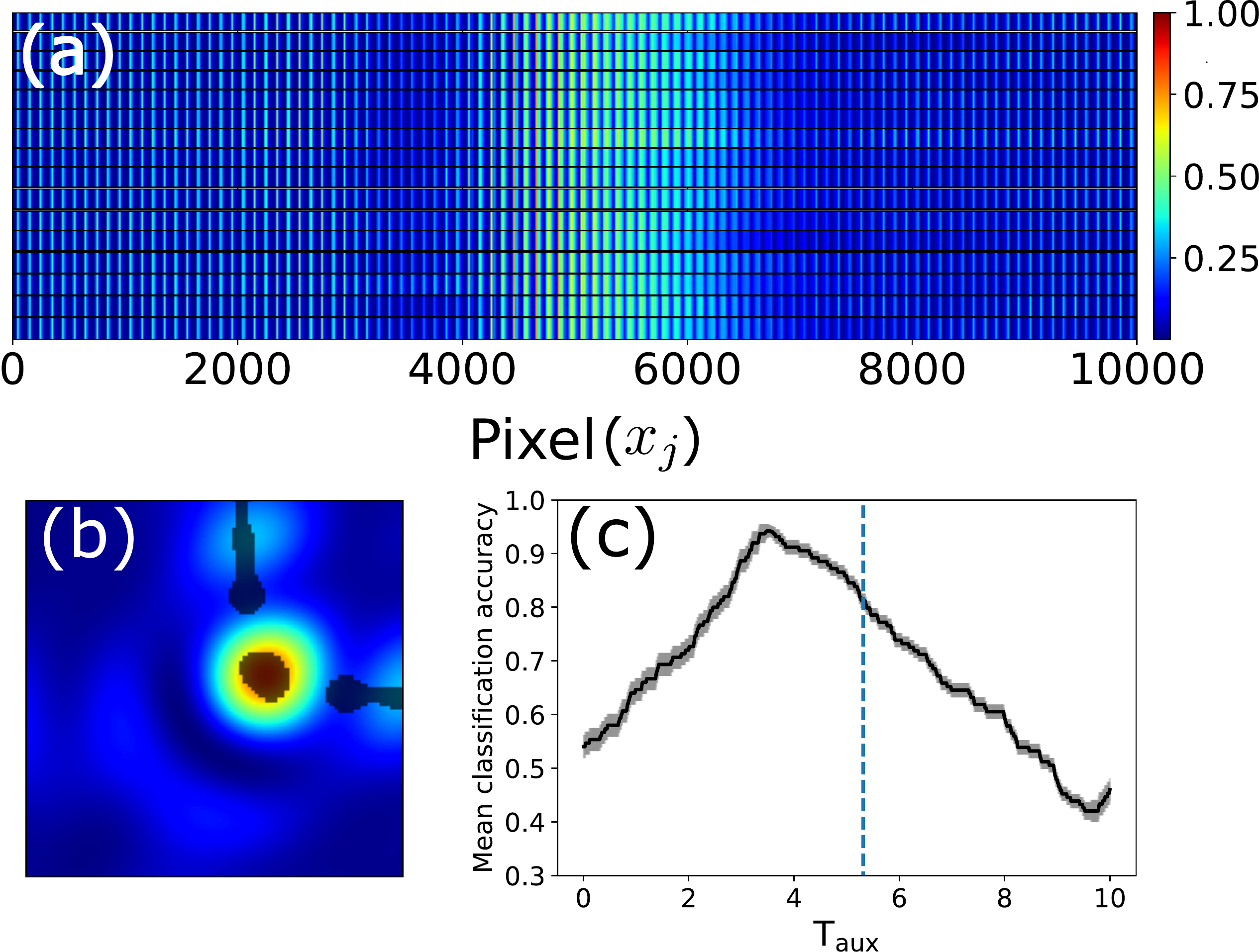}
    \caption{Analysis of the unsupervised network applied to the generalized LZ Hamiltonian. (a) Weight analysis of some of the architectures of the network, an unsymmetrical behaviour can be observed coinciding with the unsymmetrical control landscape. (b) Control landscape of the system where in black the the pixels with high importance to the network are highlighted. (c) Ensemble averaged accuracy performance function for the generalized Landau-Zener Hamiltonian with $\delta=1$, the dashed blue line marks the empirical MCT.}
    \label{fig:AC_FigureRestults}
\end{figure}

We now apply the unsupervised network to the generalized LZ Hamiltonian presented in Sec. II. In Fig. \ref{fig:AC_FigureRestults} (a) the weight analysis of the network is presented. As mentioned in Sec. II the landscapes of this control do not present symmetric topologies which is reflected in $\abs{\omega_{ij}^\nu}$. As before, we would like to relate the most important regions of the landscape to the network, thus we take all the pixels $x_j$ that satisfy the same condition used in the LZ Hamiltonian analysis but with a threshold of 0.5. The results are highlighted in black in the Fig \ref{fig:AC_FigureRestults}, where we take the total evolution time T=4 as an example. In this case, the network identifies where the region control landscape reached the global maximum and, similar to before, it also pays attention to where the sub-optimal mountains are formed thus this gives us the belief that the network is able to understand the topologies. In Sec. II we defined an empirical MCT T$_\text{MCT}$=5.31 which is compared to the MCT predicted by the ANN in Fig. \ref{fig:AC_FigureRestults} (c). Despite the fact that the prediction is lower than the empirical value (blue dashed line), we still were able to identify key factors of the networks understanding of the landscapes. We note that these results were only observed for the case of $\delta=1$, we suspect that this is due to the simplicity of the network. We understand that there are more sophisticated ANN frameworks available, such as convolution neural networks (CNN) or long short term memory networks (LSTM) which will probably yield more accurate results but as this is one of the first works covering the idea, we wanted to investigate with the simplest case we could find.

\section{Conclusion}

The external driving of quantum systems is extremely vulnerable in the presence of environments due to decoherence. A work around to this challenge is to implement fast controls, thus the knowledge of the MCT is of much importance in order to obtain revolutionary advances in quantum technologies. Despite its importance, few methods have been developed both from an analytical and numerical approach. In this paper, we propose to employ machine learning techniques to estimate the MCT of a protocol while studying what an ANN is able to learn from a system as well as its limitations. In this sense, in order to not give the network any external bias, we tested a fully unsupervised ANN scheme composed of an autoencoder and a k-means clustering method. We analysed the Landau-Zener (LZ) Hamiltonian given that it has an analytical MCT and a distinctive change in the landscape's topologies when the total evolution time is under or over the MCT. We obtained that the network is able to not only produce an estimation of the MCT but also to gain an understanding of the landscape's topologies. We moved on to implement the network in the generalized LZ Hamiltonian where similar results were yielded and also found some limitations to our very simple architecture. 

We believe this is the first approach into understand how and what the ANN does and ultimately shed light on its limitations. More work is still to be done in the future and we hope that our findings will serve as the foundations to further investigate and exploit the underling abilities of the ANN.

\begin{acknowledgements}

The authors would like to acknowledge the useful conversations with Emiliano Fortes. The work was partially supported by CONICET (PIP 11220200100568CO), UBACyT (20020130100406BA) and ANPCyT (PICT-2016-1056) 

\end{acknowledgements}

\appendix
\section{Artificial Neural Network Details} \label{AppendixANN}

In this section we will give further details as to the creation of the landscapes used as the datasets as well as the ANN used in this work. 

\begin{center}
    \begin{figure}[H]
        \centering
        \includegraphics[scale=0.5]{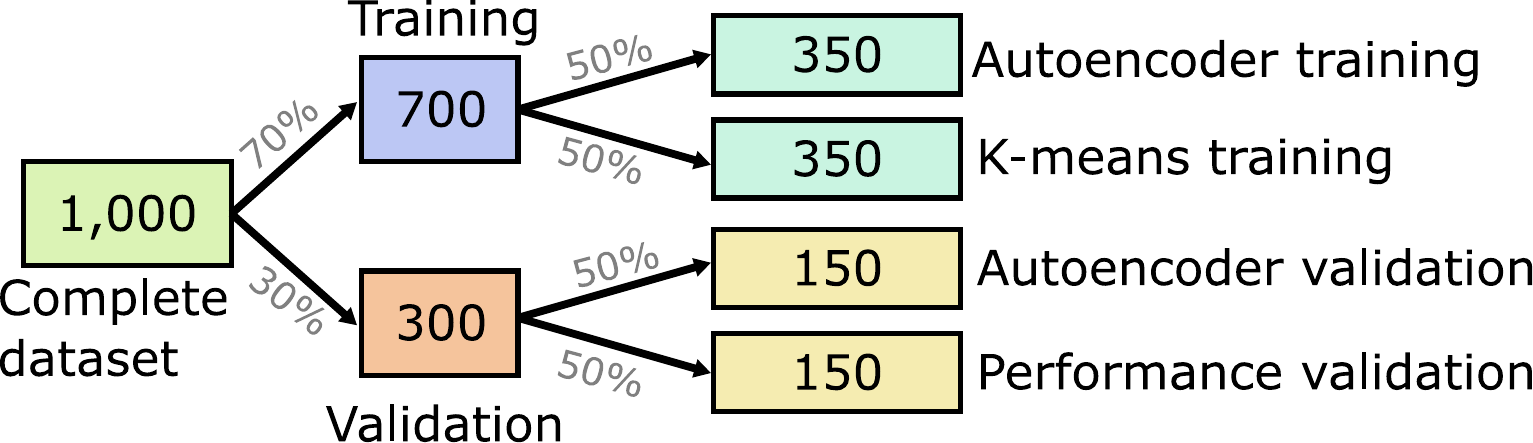}
        \caption{Illustration of the separation of the dataset for training and validation . The holdout method is implemented where 70\% of the dataset is used for training and the remaining 30\% is reserved for validation. Then the training datasets is equally separated into two parts: autoencoder training dataset and k-means training dataset. The same is  done with the validation dataset separating into the autoencoder validation dataset and performance dataset.}
        \label{fig:ANN_DATA}
    \end{figure}
\end{center}

The control landscapes used as the datasets of the ANN were constructed using the QuTiP toolbox version 4.7.0 \cite{QuTiP}. Each landscape corresponds to a certain total evolution time, T $\in$ [0.01,0.02,0.03,...,10] yielding a dataset of 1,000 components. The separation into training and validations subsets was done using the shuffled holdout method; 70\% of the data is used for training while the remaining 30\% is reserved for validation. In order to avoid a bias training in the autoencoder and k-means, the training datasets were equally separated into two parts: autoencoder training dataset and k-means training dataset. The same is  done with the validation dataset separating into the autoencoder validation dataset and performance dataset. Fig. \ref{fig:ANN_DATA} shows a visual illustration of the dataset separation to give clarity. Once the datasets are constructed, we move onto the construction of the ANN.

\begin{figure}[H]
    \centering
    \includegraphics[scale=0.6]{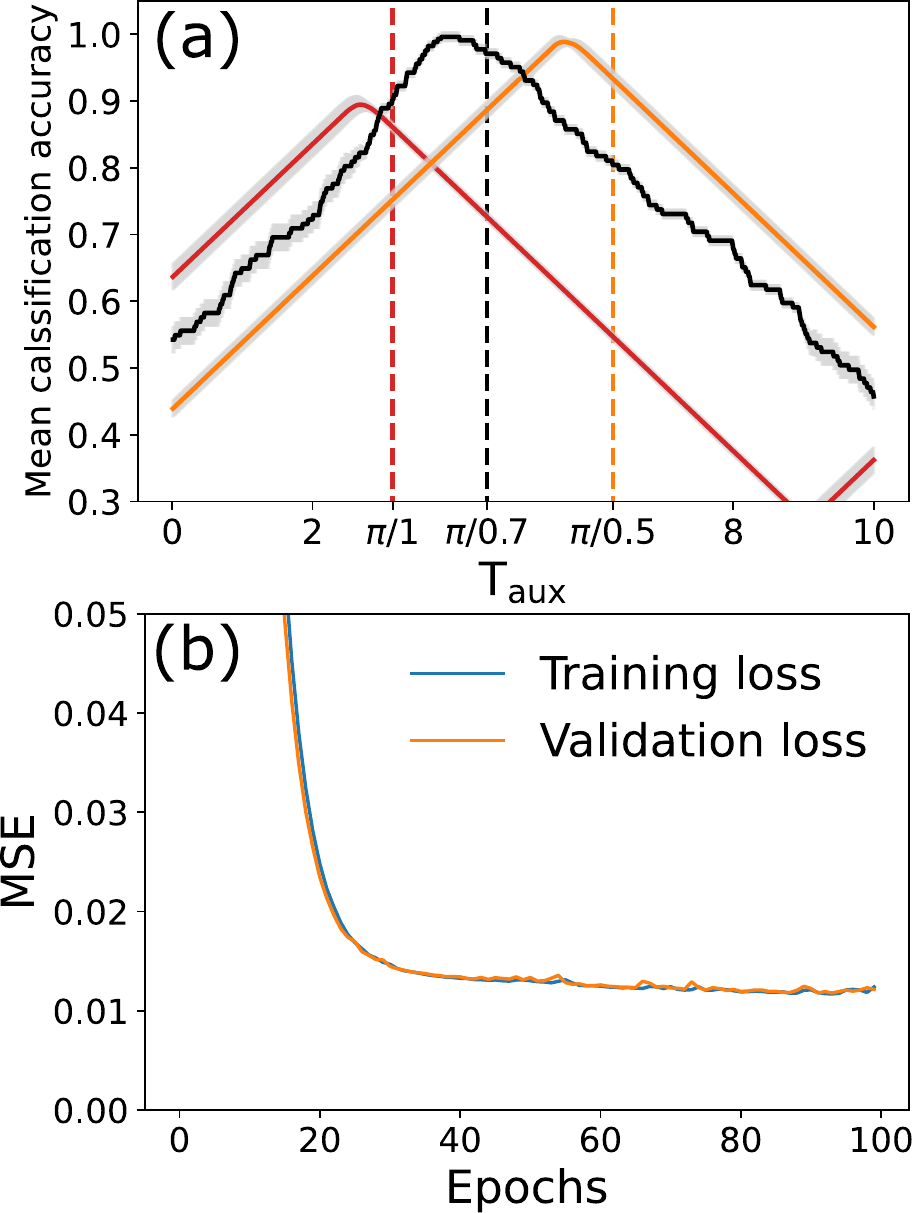}
    \caption{(a) Verification of no overfitting by training the LZ Hamiltonian (equation \ref{LZHamiltonian}) with $\delta=0.7$ (black line) and predict over the following systems; $\delta=1$ (red line) and $\delta=0.5$ (orange line). The dashed vertical lines correspond to their respective analytical MCT. (b) Training (blue) and validation (orange) loss plot for one autoencoder architecture training as a function of the number of epochs.}
    \label{fig:Appendix_Autoencoder}
\end{figure}

The autoencoder was developed using the Keras interface that runs over the Tensorflow platform version 2.6.0 \cite{Keras}. The architecture (Fig. \ref{fig:ANNArchitecture} in the main text) is composed of a flattened stage, encoder hidden layers, a layer with the smallest amount of nodes corresponding to the features, decoder hidden layers and finally a last layer that recovers the dimensions needed for the un-flattened to have the same dimensions as the input. The code allows for a deep network with many hidden layers, but for the purpose of this paper, we only required a shallow network with two encoder/decoder hidden layers. As described in the main text, ensemble averaging was implemented. For this, various architectures are to be constructed in order to form a collection of networks. This is done by building different combinations of number of nodes in the hidden layer $N_h \in$[100,110,120,...,190] and number of features extracted $f_h \in$[10,20,..,40], which yields a total of 40 architectures. The non-linear hyperbolic tangent activation function is implemented over the encoder and decoder layers in order to generate complex solutions and not restrict ourselves to Principal Component Analysis (PCA) networks.  In order to prevent overfitting while avoiding the underfitting regimen the classical L2-norm regularization term is implemented over the encoder/decoder layers with the hyperparameter $\alpha=0.005$. We show that there is no overfitting by training the LZ Hamiltonian, eq. (\ref{LZHamiltonian}), with $\delta=0.7$ and predicting over another system, for instance taking $\delta=0.5$ and $\delta=1$ [Fig. \ref{fig:Appendix_Autoencoder} (a)]. The last layer has a linear activation function and its only purpose is to recover the dimensions needed for the un-flatten layer to reshape into same dimensions as the input landscape. The training is done with the Adam stochastic gradient optimizer with 100 epochs and a batch size of 32, implementing the Mean Squared Error (MSE) loss function and a learning rate of 0.001. With these configurations, a typical training and validation plot is shown in Fig. \ref{fig:Appendix_Autoencoder} (b). In average, at the end of the training the MSE loss is of order 10$^{-2}$, this value can be lowered to 10$^{-4}$ by setting $\alpha=0$. Next we move on to explain the k-means clustering method.

The K-means methods main function is to separate a dataset into k groups. In this work we use the algorithm given by Scikit-learn version 1.0.2 \cite{scikit}. Here the unsupervised clustering is done by minimizing the inertia criterion.

\begin{figure}[H]
    \centering
    \includegraphics[scale=0.4]{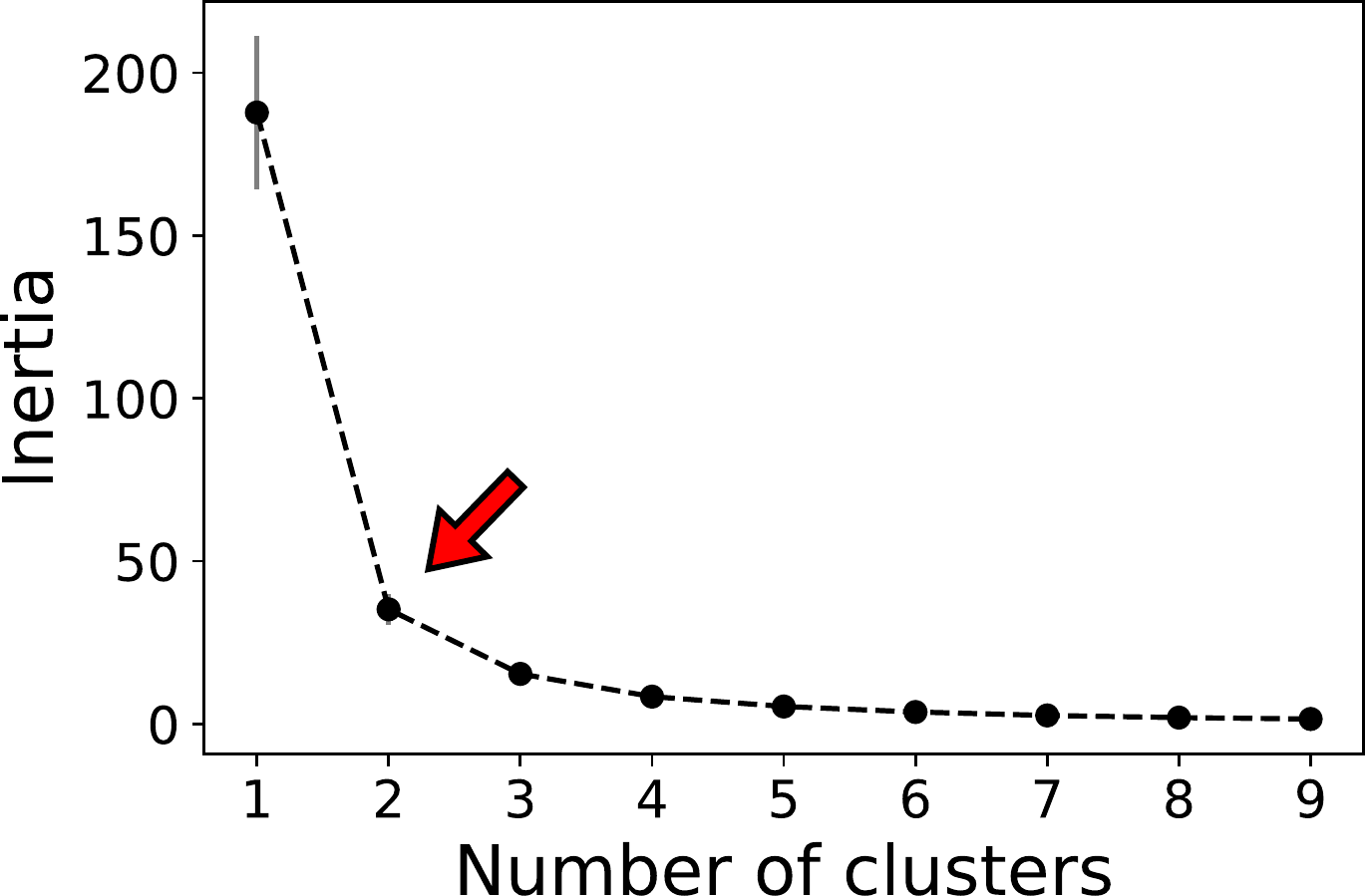}
    \caption{Elbow method in order to determine the optimal value of clusters to separate the features extracted from the landscapes. The red arrow indicates the optimized number of clusters, corresponding to k=2.}
    \label{fig:KMeans_Elbow}
\end{figure}

The features obtained by using the ANN are clustered by the unsupervised k-means. Given that the algorithm requires the number of clusters to be specified beforehand and we did not wan to force the ideal two clustering scheme, we implemented the elbow method. Fig. \ref{fig:KMeans_Elbow} shows the average result (over all the 40 architectures) of the method implemented done over the training features extracted from the LZ Hamiltonian with an energy gap $\delta=1$. As it can be seen 2 clusters is the optimized number of clusters, thus in all the networks we use the K-Means clustering into k=2 groups.

\section{Long times}\label{LongTimes}

In this section we give further details as to the minimum formed in the accuracy performance of Fig. \ref{fig:AccDelta1} in the main text. For this, we train the ANN network with the LZ Hamiltonian (equation (\ref{LZHamiltonian})), with the same details as explained in the Appendix \ref{AppendixANN}, and predict over a larger dataset that covers a total evolution time T $\in$ [0.01, 0.02, .., 49.9]. 

\begin{center}
    \begin{figure}[H]
        \centering
        \includegraphics[scale=0.8]{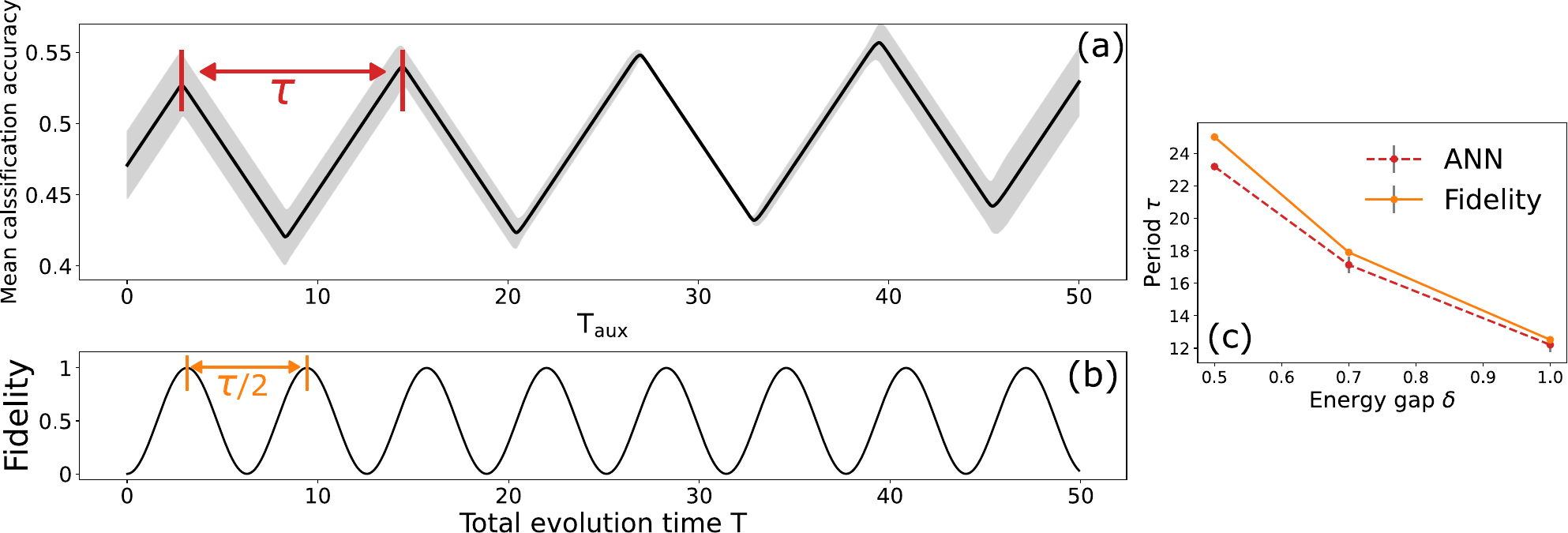}
        \caption{Analysis of prediction of ANN over long times. (a) Ensemble averaged accuracy performance function for the Landau-Zener Hamiltonian with $\delta=1$ for long times. (b) Fidelity measure in the protocol $\epsilon_1=\epsilon_2=0$ as a function of the total evolution time T. (c) The period of the oscillation in the ANN accuracy measure (red) compared with time period of the fidelity measure of the control $\epsilon_1=\epsilon_2=0$ times 2 (orange) as a function of the energy gap of the Landau-Zener Hamiltonian $\delta$.}
        \label{fig:Long_Time_Appendix}
    \end{figure}
\end{center}

In the Fig. \ref{fig:Long_Time_Appendix} (a) we show the smoothed output performance accuracy achieved for a larger time scale. The accuracy presents an oscillatory behaviour with period $\tau$. In Sec. II, we noted that the optimal control of the LZ Hamiltonian in the MCT was located in the center, that is  $\epsilon_1=\epsilon_2=0$ where, for larger times, this center maximum splitted into two symmetrical structures oscillating in time. This can be observed by analysing
F[$\epsilon_1=0,\epsilon_2=0$] (eq. \ref{Fidelity}) for different total evolution times, Fig. \ref{fig:Long_Time_Appendix} (b). Fig. \ref{fig:Long_Time_Appendix} (c) compares the period obtained in the performance of the ANN (red) with the one obtained from the fidelity measure in the center of the control landscape (orange) times 2. As one can see, for the different energy gaps $\delta$ analysed, they match pretty well. This displays that the network acquires more knowledge of landscapes than expected. Further analysis will be done in the future, given that it is outside the scope of this work.

\onecolumngrid
\twocolumngrid

\bibliography{main}

\end{document}